\documentclass[pra,aps,amsmath,amssymb,superscriptaddress,twocolumn,longbibliography]{revtex4-1}
\usepackage{graphicx,multirow}
\usepackage{color}
\usepackage{slashbox}
\usepackage{tikz}
\usepackage{verbatim}
\usetikzlibrary{arrows}

\newcommand{\halv}{\frac{1}{2}}

\newcommand{\qed}{\nobreak \ifvmode \relax \else
      \ifdim\lastskip<1.5em \hskip-\lastskip
      \hskip1.5em plus0em minus0.5em \fi \nobreak
      \vrule height0.75em width0.5em depth0.25em\fi}
\begin{document}

\title{Hierarchical axioms for quantum mechanics}   \author{S.    Aravinda}   \email{aravinda@imsc.res.in}
\affiliation{Institute  of  Mathematical  Sciences, HBNI,  C.   I.  T.
  Campus,  Taramani,  Chennai  600113,  India}  
\author{Anirban        Pathak}       \email{anirban.pathak@jiit.ac.in}
\affiliation{Jaypee   Institute  of   Information  Technology,   A-10,
  Sector-62, Noida, UP-201307, India}\author{R.   Srikanth}
\email{srik@poornaprajna.org}  \affiliation{Poornaprajna Institute  of
  Scientific  Research,   Sadashivnagar,  Bengaluru-   560080,  India}

\begin{abstract} 
The  origin of  nonclassicality  in quantum  mechanics  (QM) has  been
investigated  recently  by  a  number   of  authors  with  a  view  to
identifying  axioms that  would  single out  quantum  mechanics as  a
special theory within  a broader framework such  as convex operational
theories.   In  these  studies,  the   axioms  tend  to  be  logically
independent in  the sense that no  specific ordering of the  axioms is
implied.  Here, we identify a  hierarchy of five nonclassical features
that separate  QM from  a classical  theory: (Q1)  Incompatibility and
indeterminism; (Q2) Contextuality; (Q3) Entanglement; (Q4) Nonlocality
and  (Q5)   Indistinguishability  of  identical  particles.    Such  a
hierarchy isn't obvious when viewed from within the quantum mechanical
framework, but,  from the perspective of  generalized probability theories
(GPTs),  the  later  axioms  can  be  regarded  as  further  structure
introduced on top of earlier axioms.  Relevant toy GPTs are introduced
at each layer when useful to illustrate the action of the nonclassical
features associated with the particular layer.
\end{abstract}
\maketitle

\section{Introduction}

What exactly makes quantum mechanics (QM) nonclassical?  This question
has  been answered  in different  ways in  quantum optics,  in quantum
information        and       the        foundations       of        QM
\cite{banerjee2018physically}. For example, in quantum optics, a state
is considered to be nonclassical if its Glauber-Sudarshan $P$ function
\cite{MW95} can not be  described as classical probability distribution
function  \cite{Lou00},   i.e.,  it  takes  negative   values.   Other
quasi-probability   distributions   that   are   used   similarly   to
characterize  nonclassical  properties  of light  include  the  Wigner
distribution $W(x,p)$ and the  Husimi $Q$ distribution \cite{L97}.  In
quantum   information  theory,   we  associate   nonclassicality  with
bi-partite   or  multi-partite   \textit{quantum}  correlations   that
correspond      to     nonlocality      \cite{Gen05},     entanglement
\cite{HHH09,VDM02},  the  weaker  condition of  non-vanishing  discord
\cite{OZ01} and Einstein-Podolsky-Rosen steering.

Quantum correlations are nonclassical when  nonlocal, in that they can
violate  Bell-type   inequalities  \cite{Bel64,CHSH}   that  classical
correlations  cannot. However,  the  converse is  not
  true as  there exists nonclassical  states which are local.  In what
  follows,  this   point  will   be  illustrated  though   a  proposed
  hierarchical   structure  of   the  axioms   of  QM.    Correlation
inequalities  for temporal  situation  can be  proposed  based on  the
assumption of realism and non-invasiveness \cite{BTC+04}.

The assumptions  behind the derivation  of a Bell-type  inequality are
localism and  realism.  A  classical theory  is necessarily  local and
realist. Consequently,  a violation  of Bell's  inequality essentially
implies  non-classicality.    Likewise,  as  a  classical   theory  is
necessarily non-contextual and realist, a violation of a contextuality
inequality  would  also  imply  non-classicality, but  as  before  the
converse is  not true. 

In the  context of multi-partite  systems, the relation  between local
properties and  its non-local nature  has been extensively  studied by
various  authors \cite{Bar07,  BLM+05,  BBL+06,  BM06, BBL+07,  Bar03,
  CDP10, JH14}.   In the inverse  direction, bounds on  nonlocality in
nonsignaling  theories   have  been   derived  by   assumptions  about
monopartite  system   properties  like  uncertainty   \cite{OW10}  and
complementarity \cite{BGG+0}.  In Ref.  \cite{BBL+07}, it's shown that
any theory which  cannot be ascribed a simplex  state space structure,
with  pure states  being  one-shot distinguishable,  has a  no-cloning
theorem.  Monopartite nonclassical systems  has been considered in the
ontological  framework by  Spekkens in  Ref.  \cite{Spe05,Spe07}.   An
argument  for   the  classicality   for  discrete   physical  theories
satisfying  an  information  theoretic  axiom  is  presented  in  Ref.
\cite{PW0}.

In this  work, we identify five  basic elements that separate  QM from
classical  physics:  (Q1)   Incompatibility  and  indeterminism;  (Q2)
Contextuality;  (Q3)  Quantum  entanglement;  (Q4)  Nonlocality;  (Q5)
Indistinguishability  of   identical  particle.    A  toy   theory  is
associated  with each  axioms, and  we list  the relevant  information
theoretical tasks that can be achievable in those theories.

In  the foundations  of  quantum mechanics,  one often studies  nonclassical
features  in the framework of generalized  probability theories (GPTs)
\cite{Bar07,Har01,Har03,Man03}, with the aim to identify the minimal set
of axioms  to guarantee  the nonclassical properties  in QM.   In this
approach,   QM,  classical   theory   and   a set of  other   nonclassical
probabilistic  theories  can  be  considered as  special  cases  in  a
framework of  GPTs.  To be
  precise, by QM we mean the operational formulation of QM in terms of
  measurements, probabilities and correlations as would be observed in
  a laboratory  experiment, and states  are considered to be the lists of  probabilities of
  outcomes, with state spaces being the convex set of such operational
  states. This  formulation avoids terminology such  as Hilbert spaces
  or   phase  that   can't  be   directly  observed.    Note  that,
  mathematically, quantum  mechanics can be viewed  as a non-classical
  probability calculus  based on a non-classical  propositional logic,
  such  that  quantum  states  are  measures  on  a  suitably  defined
  non-Boolean (non-distributive), orthocomplemented lattice.

One usually associates  nonclassicality of  QM with features  like fundamental
indeterminacy   \cite{PR94},  Heisenberg   uncertainty,  monogamy   of
nonlocal correlations  \cite{Ton06}, privacy of  nonlocal correlations
\cite{Lo},    and    the     impossibility    of    perfect    cloning
\cite{WZ82,Gis}.  However, to the best of our knowledge, till now no
  hierarchical understanding of quantum nonclassicality is known, such
  that certain features are understood to  be built on top of the 
  others.   This is  of course  because  within QM,  the features  are
  mathematically interdependent, with no obvious ordering discernible.
  Here, we  address this issue,  by presenting a hierarchy  of quantum
  features, that are inspired by GPT considerations.

 There  are  certain  details  to  which  we  will  return  elsewhere:
 underlying the hierarchy is the assumption of linearity, which is not
 specific  to  quantum  mechanics,  and exists  already  in  classical
 mechanics; the  hierarchical transition from  (Q2) to (Q3)  assumed a
 tensor product structure between the state space, which itself is not
 a nonclassical feature;  (Q3) to (Q4) must  include classical notions
 of entanglement  and quantum steering.   Let us clarify that  our aim
 here is not to provide a complete axiomatic structure, rather what we
 offer  here  is a  hierarchically  axiomatic  structure inspired  by
 quantum information theory.  The idea of the hierarchy  is that later
 axioms  sit  on  top  of   the  structure  provided  by  the  earlier
 axioms.  Such  an arrangement  is  apparently  not possible,  working
 within the framework of standard QM. Hence we do it via the framework
 of GPTs.

The rest of the paper is  organized as follows. In Section \ref{sec2},
we discuss the  axiom (Q1), its relevance in  quantum cryptography and
show that it's possible to design a scheme of quantum key distribution
in  a local  toy theory  that allows  (Q1), but  does not  allow other
axioms. Similarly,  in Setion  \ref{sec3}-\ref{sec6}, we  elucidate on
the axims (Q2)-(Q5), and the toy theories valid in each layer. Quantum
infromation processing tasks that can be  performed in a layer is also
discussed.   Finally,    the   paper    is   concluded    in   Section
\ref{conclusion}.

\section{Incompatibility}\label{sec2}
 A   fundamental   feature   of  quantum   mechanics   is   the
  incompatibility   of   two   observables,  which,   for   projective
  measurements  coincides with  non-commutativity. In  the context  of
  GPTs,  the idea  of incompatibility  is formalized  by the  smallest
  value of mixing 
or ``unsharpness'' for which   two measurements  will be  jointly
  measurable, in the sense that they  can be obtained as marginals of a
  master   observable  \cite{BHS+0}.    The  cryptographic   power  of
  incompatibility,  without  any   other  nonclassical  feature  being
  assumed,  can  be illustrated  via  the  BB84-like key  distribution
  protocol, which we call ``local key distribution'' (LKD).

LKD works as follows.   Alice and Bob each have two  copies of the key
to a  strongbox. Alice opens the  box and leaves a  random bit $\kappa
\in \{0,1\}$ for Bob to read, and  then locks the box. Bob comes along
later, open the box with his key, and receives his message.

In  the  real  world,  a   classical  implementation  of  LKD  is  not
unconditionally secure, since  classical laws do not  preclude that an
eavesdropper can break open the box,  read the secret bit $\kappa$ and
then rebuild the box, without Alice and Bob knowing about it.

%%%%%%% Introduced

Consider  the two-input-two-output  operational theory  $\mathcal{T}$,
with two  dichotomic measurements  $\mathtt{X}$ and  $\mathtt{Z}$.  The
pure states of  the theory $\mathcal{T}$, which forms  the state space
$\Sigma$, are:
\begin{eqnarray}
\psi_\mathtt{X}^+ &\equiv (1,0~|~\frac{1}{2},\frac{1}{2}) \nonumber \\ 
\psi_\mathtt{X}^- &\equiv (0,1~|~\frac{1}{2},\frac{1}{2}) \nonumber \\
\psi_\mathtt{Z}^+ &\equiv (\frac{1}{2},\frac{1}{2}~|~1,0) \nonumber \\
\psi_\mathtt{Z}^- &\equiv (\frac{1}{2},\frac{1}{2}~|~0,1). 
\label{eq:uneqcr}
\end{eqnarray}
 With  the
nonsimpliciality condition,
\begin{equation}
\halv(\psi_\mathtt{X}^++\psi_\mathtt{X}^-) = \halv(\psi_\mathtt{Z}^+ +
\psi_\mathtt{Z}^-) = (\halv,\halv \mid \halv, \halv).
\label{eq:equilcr}
\end{equation}
 $\mathcal{T}$ is nonclassical.

%%%%%%%

Now consider  the following  protocol implemented in  the nonclassical
but noncontextual  theory given by  (\ref{eq:uneqcr}).  It is  the LKD
version of  the Bennett-Brassard 1984 (BB84)  quantum key distribution
(QKD) protocol  \cite{BB84}; and  we may refer  to it  as \textbf{BB84
  LKD} \cite{AAPS14}
\begin{enumerate}
\item  Alice  randomly prepares $n$ particles  in one of the  four states
  $|\psi_X^\pm),  |\psi_Z^\pm)$  given   by  (\ref{eq:uneqcr})  by
  measuring  $X$ or  $Z$  on  each particle.   She  transmits them to Bob.
\item Bob measures the particles randomly in the basis $X$ or $Z$.  He
  notes the outcomes $\tau$.
\item  On the  key string so extracted, Alice and
  Bob publicly discuss to retain only those outcomes where their bases
  agree; this  forms their raw key  string; 
\item They agree  on certain
  coordinates and announce the outcomes  on those coordinates.  If too
  many of them are mismatched, they deem the protocol round secure and
  abort  the round.   
\item  Else Alice  and Bob  proceed to  classically extract  a secure,
  smaller key from the remaining bits. \hfill \qed
\end{enumerate}

Security of the above protocol originates from  the fact that although  Eve can deterministically
extract the  encoded bit by measurement  if she measures in  the right
basis, she will produce measurement  disturbance if she gets the basis
wrong, which can be detected in Step (3).

Suppose Eve  implements such an intercept-resend  attack, by measuring
$X$ or  $Z$ on $m$  gbits from a  total of $n$  particles transmitted.
She will be able to extract $I(A{:}E) = \frac{m}{2}$  bits of information on average.
Let $f \equiv \frac{m}{n}$.  On average,  Alice and Bob will check the
basis not used  by Eve half the  times, and half of  these times, they would
obtain the  answer not encoded  by Alice.  On the  remaining fraction,
their  measured  outcome will be  consistent  with  Alice's encoded  value.
Thus, the  error observed by  Alice and Bob  is $e =  \frac{f}{4}$, so
that    on    average   Bob    receives    $I(A:B)    =   \left(1    -
h\left[\frac{f}{4}\right]\right)$ bits of  information per transmitted
gbit.

The  protocol  can be  shown  to  be  secure  if $I(A:B)  \ge  I(A:E)$
\cite{CK78}, which in this case becomes
\begin{equation}
\left(1 - h\left[\frac{f}{4}\right]\right) \ge \frac{f}{2}
\label{eq:gbitsecu}
\end{equation}
or $f \approx  0.68$, so that the tolerable error  rate $e_{\rm max} =
0.48/4$  = 17\%.   The  probability that  Eve is  not  detected on  an
attacked particle is $\frac{3}{4}$. Therefore the probability that she
escapes detection on all the $m$  bits she attacks is $(3/4)^m$, which
falls exponentially with security parameter $m$. This exponential drop
characterizes \textit{unconditional} security.

However, the security described above is not \textit{device independent} (cf.  \cite{VV14},
and references therein).  Alice and  Bob implicitly assume in the BB84
LKD  protocol  that  the   preparation  and  measurement  devices  are
trustworthy. Now suppose that the  device has been manufactured by Eve
such that  each actual  particle is replaced  by a  clandestine random
2-bit preparation, such that when Alice measures $X$, the pre-existing
bit  is  presented,  and  similarly   for  if  she  measures  $Z$.  If
(subsequently) Bob measures  the same observable, he would obtain the same
bit  as obtained by Alice,  else an  uncorrelated bit.   This reproduces  the BB84
statistics.   It is  entirely  insecure once  Eve  learns about  their
respective bases  during their  public discussion.   This is  just the
higher-dimension attack \cite{BHK05} adapted  from BB84 to the present
protocol.   Thus, BB84  LKD is  not secure  in the  device independent
scenario.

\section{Contextuality}\label{sec3}

Now  we  consider a  nonclassical  theory  with nontrivial  congruence
structure, characterized by five observables  $\mathtt{V, W, X, Y, Z}$
and the cyclic  $R_2$-chain: $R_2(\mathtt{V,W}),  R_2(\mathtt{W,X})$,  $
R_2(\mathtt{X,Y}),   R_2(\mathtt{Y,Z}),$  $R_2(\mathtt{Z,V})$, and  every
other  pair  being  incongruent.   By the  assumption  of  tomographic
separability, an arbitrary state is assumed to be completely specified
by  the  five   fiducial  probabilities  $P_\mathtt{V},  P_\mathtt{W},
P_\mathtt{X},   P_\mathtt{Y}$  and   $P_\mathtt{Z}$,  which   in  turn
determine    $P_{\mathtt{VW}},    P_{\mathtt{WX}},   $  $P_{\mathtt{XY}},
P_{\mathtt{YZ}}$ and $P_{\mathtt{ZV}}$, assumed  to be consistent with
contextual no-signaling.  We shall refer  to this theory (fragment) as
$\mathcal{T}_{\rm KCBS}$, in view of  the work where the contextuality
of such correlations was studied \cite{KCB+08}.

Consider a state $\rho$ in this contextual theory, where all the above
pairs produce  perfectly random but anticorrelated  outcomes, i.e., 01
or 10.  We can check directly that  there is no way to assign values 0
and  1  to   these  five  observables  (indeed,  any   odd  number  of
observables) in  such a  way as to  satisfy this  requirement, because
there would be a clash of values  on at least one observable. That is,
if  $\mathtt{A}$   has  value  $\mathtt{a}$,  then   $\mathtt{B}$  has
$\overline{a}$, $\mathtt{C}$ has  $a$, $\mathtt{D}$ has $\overline{a}$
so  that   $\mathtt{E}$  has   $a$  requiring  $\mathtt{A}$   to  have
$\overline{a}$,  contrary  to  assumption.    Thus,  $\rho$  does  not
correspond to  a state that  has a JD  over the five  variables, where
they take definite values.  This is  witnessed by the violation of the
KCBS inequality
\begin{equation}
\langle  \mathtt{VW}\rangle  +   \langle  \mathtt{WX}\rangle+  \langle
\mathtt{XY}\rangle    +   \langle    \mathtt{YZ}\rangle   +    \langle
\mathtt{ZA}\rangle \ge -3,
\end{equation}
where $\mathtt{V, W, X, Y, Z} = \pm1$ \cite{KCB+08}.

Now consider  the following protocol implemented  in $\mathcal{T}_{\rm
  KCBS}$: \textbf{KCBS LKD.}
\begin{enumerate}
\item Alice prepares  $n$ particles in state $\rho$,  which she leaves
  at a pre-agreed location, \textit{after} measuring each using one of
  the five observables $\mathtt{V, W, X, Y, Z}$;
\item Later  Bob arrives  at the location  and measures  the particles
  randomly in any one of these five bases.
\item Alice  and Bob announce  their measurement bases and  throw away
  the (approximately 40\%) data corresponding to instances where their
  bases aren't either identical or juxtaposed;
\item They  publicly agree  on certain  coordinates of  particles, and
  disclose their  measurement outcomes  for these.  If  their selected
  bases  are  identical  (resp.,  juxtaposed), they  verify  that  the
  outcomes are random and  identical (resp., anti-correlated).  If too
  many of them fail this criterion, then they abort the protocol.
\item  Else Alice  and Bob  proceed  to classically  distil a  secure,
  smaller key from the remaining bits.  \hfill \qed
\end{enumerate}

In this case  note that no pre-existing record can  reproduce the KCBS
perfect correlations.  Thus, any  \textit{passive} cheat device of the
type mentioned above will fail to pass the KCBS test, giving rise to a
kind  of  device independence.   Note  that  an \textit{active}  cheat
device, meaning one  that allows memory to be carried  forward in time
(which is  a kind of local  signaling), can defeat the  protocol.  For
example, the device produces an  arbitrary output when Alice measures.
Her basis and outcome information are retained in the system's memory,
so that  if Bob  measures in  the basis  as she  did, then  the device
produces the  identical (resp.,  anticorrelated) outcome if  his basis
matches  (resp.,  is  juxtaposed  to)  hers,  and  a  random  outcome
otherwise.  Note that the memory  corresponds to a kind of signal
formally, but which is not prohibited by special relativity, since the
correlation is local.  We shall refer to this scenario  where Eve that
is  restricted   from  memory   attacks  in   a  local   protocol,  as
\textit{bounded device independence}. 

Thus,  contextuality can  provide security  to LKD  in the  restricted
device-independent  scenario  where  memorylessness  (with  regard  to
Alice's  input)   is  assumed,  but  a   nonclassical  theory  without
contextuality    lacks    security    even    in    this    memoryless
device-independent  scenario.   Note that  in  the  case where  device
independence is  based on nonlocality,  the memory of  Alice's actions
(and outcomes)  cannot be  transmitted in  light-travel time  to Bob's
spacelike  separated measurement  event,  thereby  preventing Eve  from
launching the above kind of attack. In fact, without assumptions about
device trustworthiness,  that would be  the only way to  prevent Eve's
device attack.   Thus, full device independence  requires nonlocality,
and contextuality will be insufficient. 

\section{Entanglement}\label{sec4}

Quantum  entanglement,   an  important  element  which   deviates  our
classical  world  view,  acts  as  a  resource  for  many  information
theoretic  as  well  as  many computational  advantages,  providing  a
dominent  non-classical   feature  \cite{horodecki2009quantum}.   Even
though,  as  few  results  claim, the  existence  of  entanglement  in
classical optics,  it lack  any information theoretic  advantages.  An
important task  which reveals  entanglement and  also of  necessary is
teleportation. It is been proved  that entanglement is suffiecient for
teleportation in any GPTs. The important relation revealed by Spekkens
toy model  \cite{Spe07} as  well as  the toy  model proposed  by Hardy
shows   the   difference   between   entanglement   and   non-locality
\cite{hardy1999disentangling}.  They show  that both  entanglement and
teleportation is possible in local theory, thus seperating nonlocality
from entanglement.

\section{Nonlocality}\label{sec5}

In a scheme for QKD, suppose at the end of the quantum part, 
Alice's and Bob's 
joint probability $P(ab|xy)$
is described by:
\begin{equation}
P(ab|xy) = \sum_\mu  p(\mu),P(a|x,\mu) P(b|y,\mu),
\label{eq:sep}
\end{equation}
with $p(\mu)$ being a probability distribution over parameter
$\mu$. Since this is potentially preparation information with
Eve, the condition of security in the device-independent sense
is that $P(ab|xy)$ shouldn't have the above form, i.e., should
be a nonlocal correlation \cite{CHSH}. Thus, including the axiom
of nonlocal correlations being allowed in the theory
enables DI security in the usual paradigm of cryptography.

Indeed, in any non-signaling theory, nonlocality can be the
basis for distilling shared secret randomness
\cite{MAG06}.

\section{Indistinguishability of identical particles}\label{sec6}

Our  final  axiom  is indistinguishability,  relatively  less  studied
aspect  of   quantum  nonclassicality  both  in   quantum  information
processing  and the  GPT  framework.  A  task  that separates  quantum
mechanics  from classical  mechanics, that  wouldn't be  possible even
with  our  above  axioms  (i.e., quantum  mechanics  based  purely  on
distinguishable  particles), is  boson sampling,  a task  that becomes
easy when quantum indistinguishability is included.

Boson sampling, introduced in Ref. \cite{aaronson2011computational} and experimentally realized in Refs. \cite{tillmann2013experimental, wang2018toward} and references therein, is the
task  of  exactly  or  approximately  sampling  from  the  probability
distribution of identical bosons scattered by a linear interferometer.
It's widely believed that this task is intractable in the classical world (i.e, intractable for classical computers), but can be solved efficiently in the quantum world. In other words, inclusion of (Q6) in a nonclassical theory provides us the ability of solving boson sampling problem. The  appearance  of  the  permanent   in  the  outcome  statistics  of
single-photon  measurements  makes   the  task  computationally  hard,
whereas the corresponding linear optics uses only polynomial resources
to implement  it practically.  Here, we  note that the permanent  of a
square matrix in linear algebra  is a determinant-like function of the
matrix, and a special case of \textit{immanant}, a more general matrix
function.

The reason  this is  interesting from  a computational  perspective is
that  the  probability  distribution  that boson  sampling  device  is
required to  sample from,  which as  noted above  is connected  to the
permanent of  a complex  matrix. Computing the  permanent, as  well as
approximating it within multiplicative error, are known in the general
case to be in the \#P-hard computational complexity class.  Therefore,
just   as  quantum   nonlocality   suggests  a   ``behind-the-scenes''
super-classical  communication, so  does  bosonic  sampling suggest  a
``behind-the-scenes'' super-classical computing, that in some ways
is more spectacular than the quantum speedup witnessed in Shor's
prime factorization algorithm. 
  
  \section{Conclusion}\label{conclusion}

We have proposed a hierarchy of axioms for quantum mechanics, meant to
bring out the  increasing structure in the theory as  a departure from
classical mechanics, rather  than to derive quantum  mechanics per se.
These axioms, inspired by  considerations from GPTs or  convex operational  theories, are  in their  proper order
given by: (Q1) Incompatibility (or complementarity) and indeterminism;
(Q2)  Contextuality;  (Q3)  Entanglement; (Q4)  Nonlocality  and  (Q5)
Indistinguishability   of   identical   particles.  The   axioms   are
illustrated either through a  quantum information processing task that
wouldn't be possible without it, or a task from a GPT. We hope that this
work should light on the question of what makes quantum mechanics special
to be singled out by Nature.

\begin{acknowledgements}

AP thanks Defense Research  and Development Organization (DRDO), India
for    the   support    provided    through    the   project    number
ERIPR/ER/1403163/M/01/1603.   RS  thanks   Department  of  Science  \&
Technology - Science and Engineering Research Board (DST-SERB), India,
for financial support provided through the project EMR/   2016/004019.

\end{acknowledgements}

\bibliography{ax,thesis}
\end{document}